\documentclass[aps,prl,twocolumn,groupedaddress,preprintnumbers,nofootinbib]{revtex4-1}
\usepackage[dvips]{graphicx}
\usepackage[dvipsnames]{xcolor}
\usepackage{amsmath,amssymb,slashed,hyperref,mathtools,array,braket,cancel,parskip}
\usepackage[none]{hyphenat} 
\usepackage{float}
\usepackage{tikz}
\usepackage{bbold}
\usepackage{color}
\usepackage{comment} 
\usepackage[compat=1.1.0]{tikz-feynman}
\usepackage{todonotes,hyperref}
\hypersetup{
	colorlinks = true,
	linkcolor = Mahogany,
	anchorcolor = Mahogany,
	citecolor = Mahogany,
	filecolor = Mahogany,
	urlcolor = Mahogany
}
\newcommand{\nn}{\nonumber}

\newcommand{\bb}[1]{\mathbb{#1}}
\newcommand{\cl}[1]{\mathcal{#1}}

\newcommand{\Trm}[1]{\text{Tr$_-$}\left(#1\right)}
\newcommand{\Trp}[1]{\text{Tr$_+$}\left(#1\right)}
\newcommand{\Trpm}[1]{\text{Tr$_\pm$}\left(#1\right)}
\def\prd{\ref@{Phys.~Rev.~D}}        
\newcommand{\td}[1]{
	\if\notesOn1
	\todo[inline]{#1}
	\fi
}
\DeclareSymbolFont{matha}{OML}{txmi}{m}{it}
\DeclareMathSymbol{\varv}{\mathord}{matha}{118}
\def\notesOn{1}
\tikzset{
	graviton/.style={
		double,
		decoration={snake, aspect=0.75, mirror, segment length=1.5mm},
		decorate
	}
}
\tikzfeynmanset{
	bigblob/.style={
		shape=circle,
		draw=blue,
		fill=red}
}

\usepackage{enumitem,amssymb}
\newlist{todolist}{itemize}{2}
\setlist[todolist]{label=$\square$}
\usepackage{pifont}
\begin{document}
\title{On-Shell Electric-Magnetic Duality and the Dual Graviton}

\author{Nathan Moynihan$^{1,2}$ and Jeff Murugan$^{2}$}
\email{nathantmoynihan@gmail.com, jeff.murugan@uct.ac.za}
\affiliation{$^{1}$High Energy Physics, Cosmology \& Astrophysics Theory group,\\
	$^{2}$The Laboratory for Quantum Gravity \& Strings\\\\
	Department of Mathematics and Applied Mathematics, University of Cape Town, Rondebosch, Cape Town 7700, South Africa}

\begin{abstract}
	Using on-shell amplitude methods, we explore 4-dimensional Electric-Magnetic duality and its double copy. We show explicitly that the on-shell scattering amplitudes know about `dual' photons (and dual gravitons), that the off-shell photon propagator double copies to the graviton propagator and that the magnetic part of the propagator is essential for the double copy to hold. We also show that there is an equivalent gravito-magnetic part of the graviton propagator which is essential in giving rise to solutions with either angular momentum or NUT charge. Furthermore, we comment on the so-called Weinberg paradox, which states that scattering amplitudes involving the mixing of electric and magnetic monopoles cannot be Lorentz invariant, and would seem to preclude the existence of the 't Hooft-Polyakov (topological) monopole. We trace this paradox to the magnetic part of the propagator, showing that it can be eliminated if one restricts to proper orthochronous Lorentz transformations. Finally, we compute the fully relativistic cross-section for arbitrary spin dyons using the recently formulated on-shell duality transformation and show that this is always fully Lorentz invariant. 
\end{abstract}

\maketitle
\section{Introduction}
The boostrap program of the 1960's received considerable impetus by a seminal result of Weinberg in 1965 \cite{Weinberg:1965rz}. There, by demanding only Lorentz invariance of the perturbative S-matrix,
\begin{eqnarray*}
	S = \sum_{n=0}^{\infty}\frac{(-i)^{n}}{n!}\int_{-\infty}^{\infty}dt_{1}\ldots dt_{n}\mathrm{T}\left[H'(t_{1})\cdots H'(t_{n})\right]\,,
\end{eqnarray*}
where $H'(t)$ is the interaction Hamiltonian in the interaction picture, he was able to demonstrate that Maxwell's electrodynamics and Einstein's General Relativity are the unique Lorentz-invariant theories of massless particles with spin one and two respectively. Indeed, at the time, it was anticipated that the requirement of a Lorentz-invariant S-matrix was sufficient to yield all the usual properties of a local field theory such as its Feynman rules, quantum statistics, crossing symmetry, etc. Armed with more modern on-shell technology, in this letter, we revisit Weinberg's landmark results with the aim of adding the property of {\it duality} to the list above. More specifically, we will focus on the 4-dimensional electric-magnetic duality that exchanges electric and magnetic charges. The existence of magnetic currents in the theory however leads to a paradox already documented by Weinberg in \cite{Weinberg:1965rz}, namely that perturbative scattering amplitudes involving the mixing of both magnetic and electric monopoles cannot be Lorentz invariant. In the language of field theory, this can be attributed to the presence of a Dirac string which breaks Lorentz/gauge invariance, since its orientation can be chosen arbitrarily \cite{Hagen:1965zz,Zwanziger:1970hk} and may, in principle, be resolved with a proper accounting of the contribution of the string to the associated Aharonov-Bohm phase.\\ 

Intruigingly, any attempt to write down a local quantum field theory \textit{without} a Dirac string singularity necessitates the introduction of a second four-vector potential: the dual photon \cite{Cabibbo:1962td,Zwanziger:1968rs,Zwanziger:1970hk,Brandt:1977fa}. We claim that this dual photon (and its cousin the dual graviton) is known about by the on-shell amplitudes, but is usually obscured from view. While the dual of a particle shares the same on-shell properties, interactions involving charged matter and dual-charged matter breaks the Lorentz symmetry even though both sectors interacting on their own are fully Lorentz invariant. On-shell, this lack of gauge invariance in the amplitudes can be traced back to the fact that little group invariant ratios of the proportionality factor $x$ have two solutions \cite{Caron-Huot:2018ape}, corresponding to proper and improper Lorentz transformations, as we will discuss in the next section. In order to show that this symmetry breaking is not observable, we derive the fully relativistic cross-section for the scattering of two dyons of arbitrary mass and spin, showing that Lorentz invariance is restored and reproducing several results from the literature in appropriate limits. \\

It was recently shown that these $x$-ratios can be deformed by a phase, thereby complexifying the charge and giving rise to both electric and magnetically charged particle amplitudes \cite{Huang:2019cja}. We explore this idea further, showing that this operation does in fact expose the dual photon part of the propagator, already encoded in the $x$-ratios in the sense that they contain the full tensorial structure of the off-shell propagator, which has both an electric and a magnetic component. Furthermore, it was also understood how the electric-magnetic duality and magnetic monopole solutions in the abelian gauge theory double copy to gravity \cite{Luna:2015paa,Huang:2019cja,Alawadhi:2019urr,Banerjee:2019saj}. A pertinent question is then whether or not there is a double copy of the dual photon, perhaps in the form of a dual graviton. To elucidate, we show explicitly that the off-shell graviton propagator is obtained as a double copy of the photon propagator and that including the dual photon part of propagator is essential to manifesting this. The graviton propagator we obtain contains both gravito-electric (i.e. the usual de Donder gauge propagator) and gravito-magnetic components, and it is precisely this gravito-magnetic component that gives rise to gravitational physics beyond Schwarzchild, as can be seen, for example, in the recent formulations of spinning black hole solutions from scattering amplitudes \cite{Guevara:2018wpp,Chung:2018kqs,Guevara:2019fsj,Arkani-Hamed:2019ymq,Moynihan:2019bor,Burger:2019wkq,Chung:2019yfs}, as well as the Taub-NUT solution \cite{Luna:2015paa,Huang:2019cja,Alawadhi:2019urr}.  

\section{Proportionality factors and the propagator double copy}
We are interested in constructing the on-shell scattering amplitudes for two charged particles of mass $m_1$ and $m_2$ with spin $s_1$ and $s_2$ respectively. We will use the formalism developed in Ref. \cite{Arkani-Hamed:2017jhn}, where such amplitudes are constructed directly from three-particle amplitudes according to their little group transformation properties. The three-particle amplitudes themselves are built from $SL(2,\bb{C})$ invariants, and as such we must define a basis that spans $SL(2,\bb{C})$, using objects that live there: massless spinors, Levi-Cevita symbols or contractions of momentum bi-spinors. For on-shell three-particle amplitudes where two of the particles have the same mass and one particle is a massless boson, a basis of linearly independent massless spinors cannot be constructed, since the basis spinors $(u_\alpha, v_\alpha)$ are necessarily parallel to one another. We can, however, define their constant of proportionality $x$, given by \cite{Arkani-Hamed:2017jhn}
\begin{equation}\label{xvars}
x\lambda^\alpha = \frac{\tilde{\lambda}_{\dot{\alpha}}p^{\dot{\alpha}\alpha}_1}{m_1},~~~~~\frac{\tilde{\lambda}^{\dot{\alpha}}}{x} = \frac{p^{\dot{\alpha}\alpha}_1\lambda_{\alpha}}{m_1},
\end{equation} 
where $\lambda$ here is the massless particle spinor. This $x$-factor carries the little group weight of the propagating massless particle, in that $x \longrightarrow t^{-2}x$ under the particle momentum little group transformation of the form $\lambda_q\rightarrow t\lambda_q$. For $n>3$ particle amplitudes, the $x$ factors enter as little group invariant ratios, and we can think of them as the internal particle polarization vectors contracted with one of the massive particles proper velocities $u^\mu = \frac{p^\mu}{m}$, giving
\begin{equation}\label{xfactor}
x = -\sqrt{2}(\epsilon^+\cdot u_1),~~~~~\frac{1}{x} = -\sqrt{2}(\epsilon^-\cdot u_1).
\end{equation}
In spinor helicity notation we can write $x$ as
\begin{equation}\label{xsh}
x\bra{q} = [q|u_1,~~~~~\frac{[q|}{x} = \bra{q}u_1.
\end{equation}
Dotting in $\ket{\textbf{1}}$ or $|\textbf{1}]$, we find that $x$ can be used to convert between angles and squares, in the sense that 
\begin{equation}\label{xconv}
x\braket{q\textbf{1}} = [q\textbf{1}],~~~~~x\braket{q\textbf{2}} = -[q\textbf{2}].
\end{equation}
We can write the little group invariant $x$-ratios as a trace over proper velocities \cite{Huang:2019cja}
\begin{align}\label{xratio}
\frac{x_1}{x_2} &= 2(\epsilon^+\cdot u_1)(\epsilon^-\cdot u_2) = \frac{\bra{\xi}u_1|q]\bra{q}u_2|\xi]}{\braket{q\xi}[q\xi]}\nonumber\\
&= -\frac{\Trm{\xi u_1 q u_2}}{2q\cdot\xi},
\end{align}
where
\begin{align}\label{key}
\Trpm{abcd} =&~ 2\bigg[(a\cdot b)(c\cdot d) - (a\cdot c)(b\cdot d)\nn\\&+ (a\cdot d)(b\cdot c) \pm i\epsilon(abcd)\bigg],
\end{align}
and we have adopted the notation $$\epsilon(abcd) \coloneqq \epsilon^{\mu\nu\rho\sigma}a_\mu b_\nu c_\rho d_\sigma,\text{~and~}\epsilon^{\mu\nu}(ab) \coloneqq \epsilon^{\mu\nu\rho\sigma}a_\rho b_\sigma.$$ 

It is not too difficult to see that the inverse of this ratio is given by
\begin{align}\label{xratio}
\frac{x_2}{x_1} = -\frac{\Trp{\xi u_1 q u_2}}{2q\cdot\xi}.
\end{align}
Using this result, we recognise this ratio as encoding the tensorial structure of the off-shell propagator of the internal particle, including the magnetic part \cite{Weinberg:1965rz,Colwell:2015wna,Terning:2018udc}
\begin{align}\label{xprop}
\frac{x_1}{x_2} &= u_1^\mu\left(\eta_{\mu\nu} - \frac{\xi_\mu q_\nu + q_\mu\xi_\nu}{q\cdot \xi} + i\frac{\epsilon_{\mu\nu}(\xi q)}{q\cdot\xi}\right)u_2^\nu,
\end{align}
The null vector $\xi^\mu$ reflects the gauge dependence of the propagator (in this case, the light cone gauge)\footnote{In the literature, this form of the propagator is described as being in a non-covariant gauge, since the choice of $\xi^\mu$ picks out a preferred direction that is constant in spacetime. On the other hand, there are no ghosts floating around in such gauges and so we might have expected such a gauge choice to emerge from an on-shell approach.}
In this form, we can readily identify one source of Lorentz symmetry breaking, since the Levi-Civita contraction is a pseudoscalar under Lorentz transformations: $\epsilon^\prime(abcd) = \det\Lambda\,\epsilon(abcd)$, with $\det\Lambda = \pm 1$, and the choice of sign corresponding to proper and improper transformations respectively. This implies that any mixing between electric and magnetic charges will always break Lorentz invariance unless one discards improper transformations.
 
It is precisely the magnetic part of this propagator that is expected to give rise to monopole physics. We note also that the ratio of $x$'s double copies to give gravitational physics. Since the ratio encodes the details of the off-shell propagator,  the latter must also double copy. Specifically,
\begin{equation}\label{key}
\left(\frac{x_1}{x_2}\right)^2 = u_1^\mu u_2^\nu\left(\Delta_{\mu\nu\rho\sigma}^{E} + i\Delta_{\mu\nu\rho\sigma}^{B} \right)u_1^\rho u_2^\sigma,\\
\end{equation}
where $\Delta^E$ is the standard graviton propagator in the harmonic gauge (here thought of as the `gravito-electric' part of the propagator) and $\Delta^B$ the `gravito-magnetic' propagator. The latter contains an arbitrary four vector $\xi^\mu$ and connects one anti-symmetric tensor with one symmetric. Explicitly,
\begin{align*}
\Delta_{\mu\nu\rho\sigma}^{E} &= \eta_{\mu\nu}\eta_{\rho\sigma} + \eta_{\mu\rho}\eta_{\nu\sigma} - \eta_{\mu\sigma}\eta_{\nu\rho}\\
\Delta_{\mu\nu\rho\sigma}^{B} &= \frac{\eta_{\mu\nu}\epsilon_{\rho\sigma}(\xi,q)}{q\cdot\xi} + \frac{\eta_{\rho\sigma}\epsilon_{\mu\nu}(\xi,q)}{q\cdot\xi}.
\end{align*}
From this expression, we see that it is essential that we retain the magnetic part of the photon propagator in order to double copy it to the graviton propagator, since it is precisely this part that gives rise to the last two terms in the harmonic propagator, as shown in the Appendix. This piece is usually obscured in simple electric or gravitational scattering, since in the sum of the ratio and its inverse, the magnetic parts exactly cancel
\begin{align}\label{xcancel}
\frac{x_1}{x_2} + \frac{x_2}{x_1} &= 2u_1\cdot u_2,\\
\left(\frac{x_1}{x_2}\right)^2 + \left(\frac{x_2}{x_1}\right)^2 &= 2(u_1\cdot u_2)^2.
\end{align}
In order to expose the magnetic contribution we require a transformation of this ratio, of precisely the form discovered in \cite{Huang:2019cja} for charge duality or \cite{Guevara:2018wpp} for angular-momentum. It is interesting to note that in order to derive rotating black hole solutions from these amplitudes, it is essential to include the dual graviton term, since it is this that gives rise to such solutions, e.g.
\begin{align}\label{key}
e^{i2q\cdot a}\left(\frac{x_1}{x_2}\right)^2 &\propto -(u_1\cdot u_2)q\cdot a\frac{\epsilon(\xi u_1 q u_2)}{(q\cdot\xi)}\nn\\
&= -(u_1\cdot u_2)\frac{\epsilon(a u_1 q u_2)}{(q\cdot a)},
\end{align}
which eventually leads to the Kerr solution and its various generalisations \cite{Guevara:2018wpp,Chung:2018kqs,Guevara:2019fsj,Arkani-Hamed:2019ymq,Moynihan:2019bor,Burger:2019wkq,Chung:2019yfs}.
Considering the case where $q\cdot u_i = 0$, we find that the $x$-ratios are given simply by
\begin{equation}
\left(\frac{x_1}{x_2}\right)^{\pm} = u_1\cdot u_2 \pm i\frac{\epsilon(\xi u_1 q u_2)}{(q\cdot\xi)}.
\end{equation}
This can be written in a slightly simpler form by employing the Gram determinant
\begin{align}\label{gramdet}
\epsilon(\xi u_1 q u_2)^2 &= 4!\delta^\mu_{[\alpha}\delta^\nu_{\beta}\delta^\rho_{\gamma}\delta^\sigma_{\delta]}\xi_\mu u_{1\nu}q_{\rho}u_{2\sigma}\xi^\alpha u_1^\beta q^\gamma u_{2}^\delta\nonumber\\
&= -(q\cdot \xi)^2\left((u_1\cdot u_2)^2 - (u_1\cdot u_1)(u_2\cdot u_2)\right),
\end{align}
in terms of which,
\begin{align}\label{xrap}
\frac{x_1}{x_2} &= \rho \pm \sqrt{\rho^2 - 1},\\
\frac{x_2}{x_1} &= \rho \mp \sqrt{\rho^2 - 1},
\end{align}
where we have defined $\rho \coloneqq u_1\cdot u_2 $ and used the fact that $u_1^2 = u_2^2 = 1$. Here we have made explicit the fact that there are two possible solutions for $\epsilon(\xi u_1qu_2)$ in eq. \eqref{gramdet}, reflecting the choice of proper orthochronous or improper Lorentz transformations. While we could choose the positive solution here, corresponding to proper orthochronous transformations, we choose not to, in order to show explicitly that it drops out of the observables.

\section{Dyon Amplitudes from Duality Transformations}
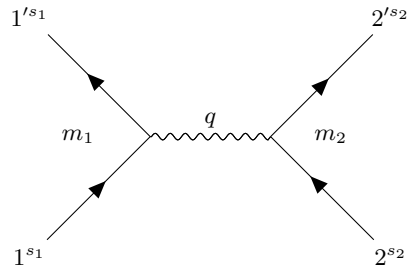
\begin{figure}[h!]
	\centering
	\begin{tikzpicture}[scale=0.8]
	\begin{feynman}  
	\vertex (a) at (-4,2) {$1'^{s_1}$};
	\vertex (b) at (-4,-2) {$1^{s_1}$};
	\vertex (c) at (2,-2) {$2^{s_2}$};
	\vertex (d) at (2,2) {$2'^{s_2}$};
	\vertex (r) at (0,0);
	\vertex (l) at (-2,0) ;
	\diagram* {
		(a) -- [anti fermion] (l) -- [photon, edge label=$q$] (r) -- [fermion] (d),
		(b) -- [fermion] (l) -- [photon] (r) -- [anti fermion] (c),
	};
	\draw (-3.2,-0.25) node[above] {$m_1$};
	\draw (1,-0.25) node[above] {$m_2$};
	\end{feynman}
	\end{tikzpicture}
	\caption{$t$-channel scattering of two particles of arbitrary spin}
	\label{electron-scattering}
\end{figure}

The orientation of the Dirac string can be attributed to the choice of vector $\xi^\mu$ in eq. \eqref{xprop} which, as we noted above, arises as the choice of gauge coming from the polarisation vector. Moreover, the propagator that emerges from the $x$-ratios is the lightcone gauge propagator, which is blind to the choice of gauge vector $\xi^\mu$. Conventionally, this is usually chosen to be a constant null vector which breaks manifest Lorentz invariance, e.g. $\xi^\mu = (1,1,0,0)$. However, when computing on-shell amplitudes, it is common to choose $\xi^\mu$ to be a dynamical gauge vector, e.g. a momentum vector of a particle attached to a different vertex, restoring manifest Lorentz invariance. In this section we will argue that while it is possible to partially restore Lorentz invariance in this way\footnote{Provided certain conditions are met.}, there will always be a source of minimal Lorentz violation. By constructing them directly on-shell, we will show that this can be seen already at the level of the amplitudes, by taking spin-$s$ electrically charged particle amplitudes and relating them to spin-$s$ dyon amplitudes, which carry both electric and magnetic charges simultaneously \cite{Weinberg:1965rz}. Towards this end, consider the $t$-channel\footnote{We don't consider the $u$ channel here since we are interested in \textit{distinguishable} dyons} charged spin-$s$ particle scattering as in Fig. \ref{electron-scattering}, where particle $1$ ($2$) has electric charge $Q_1$ ($Q_2$), mass $m_1$ ($m_2$) and spin $s_1$ ($s_2$). 

The on-shell three-particle amplitudes are given by
\begin{align}\label{3pts}
\cl{M}_3[1^s,1'^s,q^+] &= \sqrt{2}Q_1x_1\frac{\braket{\textbf{1}\textbf{1}'}^{2s_1}}{m^{2s-1}_1},\\\cl{M}_3[1,1',q^-] &= \sqrt{2}Q_1\frac1x_1\frac{[\textbf{1}\textbf{1}']^{2s_1}}{m^{2s-1}_1},
\end{align}
where here, and in what follows, $x_i$ refers to the $x$ variable of the $ii'q$ vertex. When working in the centre of mass frame, we will parametrise our external momenta as $ p_i^\mu = -(E_i,\textbf{p}),\,p_i'^\mu = (E_i,-\textbf{p})$, for $i = 1,2$ and the factor of $-1$ reflects the fact that we are working in the all-outgoing convention. With this, the energies of the particles are given by
       $E_i = E_i' = \sqrt{m_i^2 + \textbf{p}^2}$, and the usual Mandelstam variables $t = (p_1 + p_1')^2 = -q^2$ and $s = (p_1 + p_2)^2 = (E_1 + E_2)^2$. We can then construct the four-particle amplitude
\begin{align}
\cl{M}_4 
&= \frac{2Q_1Q_2\braket{\textbf{11}'}^{2s_1}\braket{\textbf{22}'}^{2s_2}}{t m_1^{2s_1 -1}m_2^{2s_2 -1}}\left(\frac{x_1}{x_2} + \frac{x_2}{x_1}\right),
\end{align}
where we have used the identity $[\textbf{11}'] = \braket{\textbf{11}'} + \frac{\bra{\textbf{1}}q|\textbf{1}']}{m_1}$ along with eq. \eqref{xconv}.
This amplitude represents the scattering of two electrically charged particles of spin $s_1$ and $s_2$. In order to endow these particles with a magnetic charge, turning them into dyons, we will utilise the electric-magnetic duality transformation \cite{Huang:2019cja} given by
\begin{equation}\label{duality-xf}
x_i\longrightarrow x_i e^{i\theta_i}.
\end{equation}
As we have seen, this transformation precisely exposes the magnetic part of the propagator, which would otherwise have cancelled out due to eq. \eqref{xcancel}. The electric charge $e_i$ is now given by the real part, and the magnetic charge $g_i$ by the imaginary part, i.e.
\begin{equation}\label{dyoncharge}
Q_je^{i\theta_j} = e_j + ig_j.
\end{equation}
Consequently, the amplitude for two dyons of spin $s_1$ and $s_2$ is given by
\begin{align}
\cl{M}_4 &= \frac{2Q_1Q_2\braket{\textbf{11}'}^{2s_1}\braket{\textbf{22}'}^{2s_2}}{t m_1^{2s_1 -1}m_2^{2s_2 -1}}\left(\frac{x_1}{x_2}e^{i(\theta_1 - \theta_2)} + \frac{x_2}{x_1}e^{-i(\theta_1 - \theta_2)}\right)\nn\\
&= \frac{4\braket{\textbf{11}'}^{2s_1}\braket{\textbf{22}'}^{2s_2}}{t m_1^{2s_1 -1}m_2^{2s_2 -1}}\nn\\&~~~~~\times\left(\rho(e_1e_2 + g_1g_2) \mp i\sqrt{\rho^2 - 1}(e_1g_2 - e_2g_1) \right).
\end{align}
%

It is interesting that the only source of Lorentz violation in this amplitude is the choice of sign, rather than the presence of a constant (unphysical) vector which breaks spacetime isometry. We may eventually wish to take the non-relativistic limit where $\rho \longrightarrow 1$, however before we do this we need to identify the relevant terms that contribute to electric/magnetic mixing. To facilitate the comparison, and to make contact with the literature, let's define the constant vectors $\textbf{k}_i = (e_i,g_i)$ in charge space. In order to write down a covariant expression with `more' Lorentz symmetry, we instead define a \textit{dynamical} four-vector $k_i^\mu$, requiring that the following conditions are met
\begin{equation}\label{key}
k_i\cdot u_i = 0,~~~~~k_1\cdot k_2 = e_1e_2 + g_1g_2.
\end{equation}
If these are satisfied, we can write
\begin{equation}
e_1 g_2 - e_2 g_1 = \frac{\epsilon(u_1,u_2,k_1,k_2)}{\sqrt{\rho^2 - 1}}, 
\end{equation}
in terms of which the amplitude reads
\begin{equation}
\cl{M}_4 = \frac{4\braket{\textbf{11}'}^{2s_1}\braket{\textbf{22}'}^{2s_2}}{t m_1^{2s_1 -1}m_2^{2s_2 -1}}\left(\rho (k_1\cdot k_2) \mp i\epsilon(u_1,u_2,k_1,k_2)\right).
\end{equation}
If the above conditions can be met, then the Lorentz symmetry breaking here is minimal, in the sense that one simply has to choose between proper and improper Lorentz transformations. Despite this amplitude not being {\it strictly} Lorentz invariant, it's square always will be, since
\begin{align}\label{key}
|\cl{M}_4| &\propto \rho^2(\textbf{k}_1\cdot \textbf{k}_2)^2 + \det(G(u_1,u_2,k_1,k_2))
\end{align}

With this in mind, we now move on to compute the cross section in the centre of mass (COM) frame, choosing our particles to be aligned along the $z$ direction so that $k_i^\mu = (0,e_i,g_i,0)$ and
\begin{equation}
u_1^\mu = \frac{1}{m_1}(E_1,0,0,p),~~~~~u_2^\mu = \frac{1}{m_2}(E_2,0,0,-p)
\end{equation}
In this particular frame, we can write
\begin{equation}
\epsilon(u_1,u_2,k_1,k_2) = \frac{(E_1 + E_2)|p|}{m_1m_2}(\textbf{k}_1\times \textbf{k}_2) = \frac{\sqrt{s}|p|}{m_1m_2}(\textbf{k}_1\times \textbf{k}_2),
\end{equation}
and we find the amplitude
\begin{equation}
\cl{M}_4 = \frac{4\braket{\textbf{11}'}^{2s_1}\braket{\textbf{22}'}^{2s_2}}{t m_1^{2s_1 -1}m_2^{2s_2 -1}}\left(\rho (\textbf{k}_1\cdot \textbf{k}_2) \mp i\frac{\sqrt{s}|p|}{m_1m_2}(\textbf{k}_1\times \textbf{k}_2)\right).
\end{equation}
The cross-section in the COM frame is given by\footnote{This is related to $\frac{d\sigma}{d\Omega}$ by $\pi dt = |\textbf{p}_1||\textbf{p}_1'| d\Omega$.}
\begin{equation}
\frac{d\sigma}{dt} = \frac{1}{64\pi s |\textbf{p}_{1}|^2}|\cl{M}_4|^2,
\end{equation}
where the square modulus of the amplitude is
\begin{align}
|\cl{M}_4|^2 &= \frac{16(2p_1\cdot p_1')^{2s_1}(2p_2\cdot p_2')^{2s_2}}{t^2(m_1^{2s_1 -1}m_2^{2s_2 -1})^2}\nn\\&\times\bigg[\rho^2(\textbf{k}_1\cdot \textbf{k}_2)^2 + \frac{s|p|^2}{m_1^2m_2^2}(\textbf{k}_1\times \textbf{k}_2)^2\bigg].
\end{align}
The cross section is then given by
\begin{align}
\frac{d\sigma}{dt} &= \frac{(t - 2m_1^2)^{2s_1}(t-2m_2^2)^{2s_2}}{4\pi t^2 s|p|^2(m_1^{2s_1 -1}m_2^{2s_2 -1})^2}\\&\times\bigg[\frac{(s - m_1^2 - m_2^2)^2}{4m_1^2m_2^2}(\textbf{k}_1\cdot \textbf{k}_2)^2 + \frac{s|p|^2}{m_1^2m_2^2}(\textbf{k}_1\times \textbf{k}_2)^2\bigg],\nn
\end{align}
where we have used $\rho = u_1\cdot u_2 = \frac{E_1E_2 + |p|^2}{m_1m_2} = \frac{s-m_1^2-m_2^2}{2m_1m_2}$. 
Note also that for $t\ll m_i^2$ the kinematic spin dependence drops out of the cross section and we have
\begin{align}
\frac{d\sigma}{dt}\Bigg|_{t\ll m^2} &= (-2)^{2(s_1 + s_2)}\frac{1}{4\pi t^2 s|p|^2}\\
&\times\bigg[\frac{(s - m_1^2 - m_2^2)^2}{4m_1m_2}(\textbf{k}_1\cdot \textbf{k}_2)^2 + \frac{s|p|^2}{m_1m_2}(\textbf{k}_1\times \textbf{k}_2)^2\bigg].\nn
\end{align} 
In the high energy limit, we have that $s \gg m_i^2$ and $|p| = \frac{\sqrt{s}}{2}$ such that
\begin{align}
\frac{d\sigma}{dt}\Bigg|_{s\gg m^2} &= \frac{(t - 2m_1^2)^{2s_1}(t-2m_2^2)^{2s_2}}{4\pi t^2(m_1^{2s_1 }m_2^{2s_2})^2}\nn\\&~~~~~\times\bigg[(\textbf{k}_1\cdot \textbf{k}_2)^2 + (\textbf{k}_1\times \textbf{k}_2)^2\bigg],
\end{align}
Choosing $s_1 = s_2 = 0$ exactly reproduces the result derived for scalar dyons in the non-perturbative Eikonal limit \cite{Gamberg:1999hq}. We are also free to take the non-relativistic limit, in which $s\rightarrow (m_1+m_2)^2$ for small $|p|$. This gives the cross section
\begin{align}
\frac{d\sigma}{dt}\Bigg|_{s\simeq m^2} &= \frac{(t - 2m_1^2)^{2s_1}(t-2m_2^2)^{2s_2}}{4\pi t^2(m_1^{2s_1 -1}m_2^{2s_2 -1})^2}\nn\\&\times\bigg[\frac{(\textbf{k}_1\cdot \textbf{k}_2)^2}{(m_1+m_2)^2|p|^2} + \frac{(\textbf{k}_1\times \textbf{k}_2)^2}{m_1^2m_2^2}\bigg]
\end{align}
We can again look at the $s_1 = s_2 = 0$ case for comparison with the literature, which gives
\begin{equation}
\frac{d\sigma}{dt}\Bigg|_{s\simeq m^2} = \frac{1}{4\pi t^2}\bigg[\frac{\mu^2}{|p|^2}(\textbf{k}_1\cdot \textbf{k}_2)^2 + (\textbf{k}_1\times \textbf{k}_2)^2\bigg],
\end{equation}
where $\mu = \frac{m_1m_2}{m_1 + m_2}$ is the reduced mass. Taking $|p|^2 = \mu^2 \textbf{v}^2$ and $t = 4E^2\sin^2\theta/2 = \mu^2 \textbf{v}^2\sin^2\theta/2$, reduces this to
\begin{equation}
\frac{d\sigma}{dt}\Bigg|_{s\simeq m^2} = \frac{1}{4\pi \mu^4 \textbf{v}^4}\bigg[\frac{(\textbf{k}_1\cdot \textbf{k}_2)^2}{\textbf{v}^2} + (\textbf{k}_1\times \textbf{k}_2)^2\bigg]\frac{1}{\sin^4(\theta/2)},
\end{equation}
which agrees with the previously derived result \cite{Schwinger:1976fr} in the small angle limit.

\section{Conclusion}
In this paper, we have further explored the recent on-shell formulation of 4-dimensional electric-magnetic duality and its double copy. More precisely, we have shown that the on-shell amplitudes of massive particles exchanging massless particles of helicity $h = (\pm1,\pm2)$ contain propagators that know of both photons (gravitons) and dual photons (dual gravitons). Curiously, for the double copy to hold, it is essential that the dual photons are included in the single copy, because without them the tensorial structure of the propagator is simply, and incorrectly, $\Delta_{\mu\nu\rho\sigma} = \eta_{\mu\nu}\eta_{\rho\sigma}$. The full de Donder gauge propagator is only recovered from the double copy of the dual photon, which gives the other two terms, along with terms involving the dual graviton. Taking the double copy seriously, we see that it does in fact predict the existence of such a dual graviton (and, by extension, of an S-dual theory of gravity), as predicted in many models of string theory \cite{West:2002jj,Hull:2000zn,Hull:2001iu,Nicolai:2005su} and a subject which has been intensely studied in its own right \cite{Deser:1983wb,GarciaCompean:1998qh,Nieto:1999pn,Bekaert:2002uh,Shen:2003sd,Henneaux:2004jw,Bunster:2006rt}. When considering rotating black holes, we argued that the angular momentum is found by exposing the dual graviton part of the propagator, and that without this no Kerr or Kerr-like solution would be found from the amplitudes. Specifically, we found that exponential transformations of the on-shell three-particle amplitudes, whether of the soft factors \cite{Guevara:2018wpp} or a pure phase \cite{Huang:2019cja}, exposed the dual photon or dual graviton contributions of the propagator that are usually obscured. In many ways, electric-magnetic duality, innocuous as it might seem, was the seed for many of the remarkable developements in mathematical physics over the past half century, from the discovery of Seiberg duality in supersymmetric quantum field theory to the recent progress in the geometric Langlands program \cite{Kapustin:2006pk}. Most of these dicoveries were made with the use of conventional quantum field theory tools of Feynman diagrams and path integrals. However, as the surge of activity in the amplitudes program has demonstrated, modern on-shell methods constitute an entirely new way of thinking about quantum field theory and gravity. Taken together with other remarkable developments like the numerical and conformal boostrap programs and the Ryu-Takayanagi formula for entanglement entropy, it is clear that we stand on the precipice of an exciting new era in theoretical physics.

\section{Acknowledgements}
We thank Daan Burger for many insightful discussions and collaboration on related topics. JM is supported by the NRF of South Africa under grant CSUR 114599. NM is supported by funding from the DST/NRF SARChI in Physical Cosmology.  Any opinion, finding and conclusion or recommendation expressed in this material is that of the authors and the NRF does not accept any liability in this regard.
\appendix
\begin{widetext}
\section{Appendix: Graviton Propagator as a Double Copy}\label{app1}
In this section we give some details regarding the double copy of the propagator. To begin with, we will derive the propagators for the photon and its dual, then show explicitly that this is encoded in the $x$-ratios, and that it double copies to the graviton propagator with dual-graviton (gravito-magnetic) piece.

Massless photon polarization vectors $\epsilon_\mu^h(q)$, where $h$ is the helicity, satisfy $q\cdot \epsilon^h(q) = 0$ and $\epsilon^h(q)\cdot(\epsilon^{h'}(q))^* = -\delta_{hh'}$. Dual photon polarization vectors are defined via the dual Faraday tensor
\begin{equation}\label{key}
\tilde{F}^{\mu\nu} = \partial^{[\mu}\tilde{A}^{\nu]} = \epsilon^{\mu\nu\rho\sigma}F_{\rho\sigma},
\end{equation}
from which we can easily derive their relation to the standard photon polarizations (in momentum space) via
\begin{equation}\label{dualpol}
\tilde{\epsilon}_h^\mu = i\frac{\epsilon^{\mu\nu\rho\sigma}\xi_\nu q_\rho\epsilon_{h,\sigma}}{q\cdot\xi},
\end{equation}
where $\xi$ is an arbitrary vector that must satisfy $\xi\cdot\epsilon(q) = 0$ and $\xi\cdot q \neq 0$. The standard sum over helicities of the usual polarization is well known, and in the light-cone gauge, is given by
\begin{equation}\label{key}
\sum_h\epsilon_\mu^h(q)(\epsilon_\mu^h(q))^* = -\eta_{\mu\nu} + \frac{\xi_\mu q_\nu + q_\mu\xi_\nu}{q\cdot \xi},
\end{equation}
In order to include the dual photon, we must sum over its states too, meaning we need to evaluate
\begin{equation}\label{fullprop}
\Delta_{\mu\nu}^\pm = \frac12\sum_h\epsilon_\mu^h(q)(\epsilon_\nu^h(q))^* - \frac12\sum_h\tilde{\epsilon}_\mu^h(q)(\tilde{\epsilon}_\nu^h(q))^* \pm \sum_h\tilde{\epsilon}_\mu^h(q)(\epsilon_\nu^h(q))^*.
\end{equation}
This is due to the fact that the free Lagrangian for dual spin-1 particles is usually formulated as
\begin{equation}\label{key}
\cl{L} = \frac14(F^{\mu\nu} \pm i \tilde{F}^{\mu\nu})^2,
\end{equation}
which accounts for the factor of $i$ in eq. \eqref{dualpol} and for the $\pm$ in eq. \eqref{fullprop}.

The second term is equivalent to the first, up to a minus sign, as expected by the duality. This can be seen by considering
\begin{equation}\label{key}
\frac12\sum_h\tilde{\epsilon}_\mu^h(q)(\tilde{\epsilon}_\nu^h(q))^* = -\frac12\frac{\epsilon_{\mu\rho\sigma\lambda}\xi^\rho q^\sigma\epsilon_{\nu\alpha\beta\delta}\xi^\alpha q^\beta}{(q\cdot\xi)^2}\sum_h\epsilon^\lambda_h(q)(\epsilon^\delta_h(q))^* = -\frac12\left(-\eta_{\mu\nu} + \frac{\xi_\mu q_\nu + q_\mu\xi_\nu}{q\cdot \xi}\right).
\end{equation}
The third term is more interesting, since it somehow encodes both electric and magnetic information simultaneously. Evaluating this, we find that it gives
\begin{equation}\label{key}
\sum_h\tilde{\epsilon}_\mu^h(q)(\epsilon_\nu^h(q))^* = i\frac{\epsilon_{\mu\rho\sigma\lambda}\xi^\rho q^\sigma}{q\cdot\xi}\sum_h\epsilon^\lambda_h(q)(\epsilon_{\nu,h}(q))^* = i\frac{\epsilon_{\mu\nu\rho\sigma}\xi^\rho q^\sigma}{q\cdot\xi}
\end{equation}
and thus we find that the full propagator is 
\begin{equation}\label{key}
\Delta_{\mu\nu}^\pm = -\eta_{\mu\nu} + \frac{\xi_\mu q_\nu + q_\mu\xi_\nu}{q\cdot \xi} \pm i\frac{\epsilon_{\mu\nu}(\xi q)}{q\cdot\xi},
\end{equation}
which matches the propagator in eq. \eqref{xprop} that we found encoded in the on-shell amplitudes. We see then that what we usually evaluate as the propagator is actually formulated as
\begin{equation}\label{key}
\frac12\left(\Delta_{\mu\nu}^+ + \Delta_{\mu\nu}^-\right) = -\eta_{\mu\nu} + \frac{\xi_\mu q_\nu + q_\mu\xi_\nu}{q\cdot \xi}
\end{equation}
and that a consistent way to expose the dual photon propagator is to transform $\Delta_{\mu\nu}^\pm \rightarrow e^{\pm if}\Delta_{\mu\nu}^\pm$, where $f$ is any function.

Finally, we consider the double copy, in which the propagator is encoded into the $x$-ratio via
\begin{equation}\label{key}
\left(\frac{x_1}{x_2}\right)^2 = u_1^\mu u_2^\nu \Delta_{\mu\nu\rho\sigma}^+u_1^\rho u_2^\sigma = u_1^\mu u_2^\nu \Delta_{\mu\nu}^+\Delta_{\rho\sigma}^+u_1^\rho u_2^\sigma,
\end{equation}
We can therefore determine $\Delta_{\mu\nu\rho\sigma}^\pm$, where we will ignore any terms that vanish by $q\cdot u_i = 0$, giving 
\begin{align}\label{key}
\Delta_{\mu\nu\rho\sigma}^\pm &= \left(\eta_{\mu\nu} - \frac{\xi_\mu q_\nu + q_\mu\xi_\nu}{q\cdot \xi} \pm i\frac{\epsilon_{\mu\nu}(\xi, q)}{q\cdot\xi}\right)\left(\eta_{\rho\sigma} - \frac{\xi_\rho q_\sigma + q_\rho\xi_\sigma}{q\cdot \xi} \pm i\frac{\epsilon_{\rho\sigma}(\xi, q)}{q\cdot\xi}\right)
\\&= \eta_{\mu\nu}\eta_{\rho\sigma} - \frac{\epsilon_{\mu\nu}(\xi,q)\epsilon_{\rho\sigma}(\xi,q)}{(q\cdot\xi)^2} \pm i\frac{\eta_{\mu\nu}\epsilon_{\rho\sigma}(\xi,q)}{q\cdot\xi} \pm i\frac{\eta_{\rho\sigma}\epsilon_{\mu\nu}(\xi,q)}{q\cdot\xi}\\
&= \eta_{\mu\nu}\eta_{\rho\sigma} + \eta_{\mu\rho}\eta_{\nu\sigma} - \eta_{\mu\sigma}\eta_{\nu\rho} \pm i\frac{\eta_{\mu\nu}\epsilon_{\rho\sigma}(\xi,q)}{q\cdot\xi} \pm i\frac{\eta_{\rho\sigma}\epsilon_{\mu\nu}(\xi,q)}{q\cdot\xi}\\
&= \Delta_{\mu\nu\rho\sigma}^{E} \pm i\Delta_{\mu\nu\rho\sigma}^{B}
\end{align}
where we have used the fact that
\begin{align}\label{key}
\epsilon_{\mu\nu}(\xi,q)\epsilon_{\rho\sigma}(\xi,q) &= \eta_{\lambda\mu}\eta_{\chi\nu}\epsilon^{\lambda\chi}(\xi,q)\epsilon_{\rho\sigma}(\xi,q) = 4!\eta_{\lambda\mu}\eta_{\chi\nu}\delta^\lambda_{[\rho}\delta^\chi_\sigma\delta^\alpha_\gamma\delta^\beta_{\delta]}\xi^\gamma\xi_\alpha q^\delta q_\beta\\
&= -(q\cdot\xi)^2\left(\eta_{\mu\rho}\eta_{\nu\sigma} - \eta_{\mu\sigma}\eta_{\nu\rho}\right).
\end{align}
We recognise $\Delta_{\mu\nu\rho\sigma}^{E}$ as being the usual de-Donder gauge graviton propagator, and $\Delta_{\mu\nu\rho\sigma}^{B}$ as a dual graviton contribution, which gives rise to physics beyond Schwarzchild (or Reissner-Nordstrom) in the gravitational sector. We see then that this is exactly the part of the propagator that is exposed by the duality transformation proposed in Ref. \cite{Huang:2019cja}, which would otherwise vanish in the sum of $x$-ratios. We observe that the Lorentz violation discussed in section 1 also double copies, and is a reflection of the so-called wire singularities usually found in the Taub-NUT metric.

Taking the double copy seriously, we also see that it predicts a dual graviton of the form
\begin{equation}\label{key}
\tilde{h}^{\mu\nu}(q) = i\frac{\epsilon^{\mu\rho\sigma\tau}\xi_\rho q_\sigma h_{\tau}^\nu}{q\cdot\xi},
\end{equation}
which does indeed give rise to the full graviton/dual-graviton propagator if we consider the same procedure as for the photon. This relationship can be easily derived by considering the dual Riemann tensor contracted with two arbitrary vectors that satisfy $\xi\cdot h^\mu = 0$, i.e. $\xi^\mu\xi^\rho \tilde{R}_{\mu\nu\rho\sigma}$ and comparing the linearized dual $\tilde{h}_{\mu\nu}$ with the graviton, where 
\begin{equation}\label{key}
\tilde{R}_{\mu\nu\rho\sigma} = \frac12\epsilon_{\mu\nu\alpha\beta}R^{\alpha\beta}_{\rho\sigma}.
\end{equation}
We leave the intriguing study of such a dual graviton to the future.
\end{widetext}
\bibliography{dyons}
\end{document}